\documentclass[letterpaper]{article}
\usepackage{natbib,alifeconf}

\title{Exploring the coevolution of predator and prey morphology and behavior}
\author{Randal S.~Olson$^{1,3}$, Arend Hintze$^{2,3}$, Fred C.~Dyer$^{2,3}$, Jason H.~Moore$^{1}$,~\and Christoph Adami$^{2,3}$\\
\mbox{}\\
$^1$University of Pennsylvania, Philadelphia, PA 19143\\
$^2$Michigan State University, East Lansing, MI 48824\\
$^3$BEACON Center for the Study of Evolution in Action, East Lansing, MI 48824\\
olsonran@upenn.edu, hintze@msu.edu, fcdyer@msu.edu, jhmoore@upenn.edu, adami@msu.edu\\
}

\newlength{\HalfPage}
\begin{document}
\maketitle

\begin{abstract}A common idiom in biology education states, ``Eyes in the front, the animal hunts. Eyes on the side, the animal hides.'' In this paper, we explore one possible explanation for why predators tend to have forward-facing, high-acuity visual systems. We do so using an agent-based computational model of evolution, where predators and prey interact and adapt their behavior and morphology to one another over successive generations of evolution. In this model, we observe a coevolutionary cycle between prey swarming behavior and the predator's visual system, where the predator and prey continually adapt their visual system and behavior, respectively, over evolutionary time in reaction to one another due to the well-known ``predator confusion effect.'' Furthermore, we provide evidence that the predator visual system is what drives this coevolutionary cycle, and suggest that the cycle could be closed if the predator evolves a hybrid visual system capable of narrow, high-acuity vision for tracking prey as well as broad, coarse vision for prey discovery. Thus, the conflicting demands imposed on a predator's visual system by the predator confusion effect could have led to the evolution of complex eyes in many predators.
\end{abstract}
\hspace{0.1in}

{\parindent0pt Keywords: {\it swarming behavior}, {\it predator confusion effect}, {\it predator-prey coevolution}, {\it visual acuity}}

\hspace{0.1in}

\section*{Introduction}

``Eyes in the front, the animal hunts. Eyes on the side, the animal hides.'' So goes the common idiom in biology education when teaching students how to classify animal skulls. It is widely believed that forward-facing, high-acuity visual systems play an important role in predation, for example, in dragonflies catching flying prey~\citep{Olberg2012}. Despite this common observation, we have little empirical evidence explaining the evolutionary history of these focused visual systems observed in so many predators. In this paper, we explore one hypothesis for why predators tend to evolve complex visual systems: the conflicting demands imposed on a predator's visual system by the well-known ``predator confusion effect'' could have led to the evolution of complex eyes in many predators.

In previous work, we have shown that not all prey evolve to respond to the presence of predators by hiding or fleeing~\citep{Olson2013PredatorConfusion,Olson2013SelfishHerd,Haley2014,Haley2015,Olson2015ManyEyes,Olson2016SelfishHerd}. In fact, some prey species have evolved to stay together, form swarms, and defend themselves as a group for a variety of hypothesized reasons~\citep{Krause2002}. For example, swarming is hypothesized to improve group vigilance~\citep{Pulliam1973,Treisman1975,Kenward1978,Treherne1981}, reduce the chance of being encountered by predators~\citep{Treisman1975,Inman1987}, dilute an individual's risk of being attacked~\citep{Hamilton1971,Foster1981,Treherne1982}, and reduce predator attack efficiency by confusing the predator, i.e., the predator confusion effect~\citep{Jeschke2007,Ioannou2008}. As such, swarming opens the possibility for an evolutionary ``arms race'' between predators and their prey~\citep{Vermeij1987}, where the predator and prey continually adapt to one another over many generations of evolution.

Here we use an agent-based computational model of evolution to study the coevolutionary dynamics between predator and prey~\citep{Olson2015Thesis}. We implement the predator confusion effect as a simple perceptual constraint on the predator's visual system, and allow both the predator and prey behavior to coevolve over successive generations of evolution (as in~\citealt{Olson2013PredatorConfusion}). In addition, we allow the predator visual system to simultaneously evolve, which enables us to explore how the predator visual system adapts in response to the prey behavior. From these experiments, we discover a coevolutionary cycle between prey swarming behavior and the predator's visual system. From further analysis, we discover that the predator visual system is the primary driver of this cycle: when the predator evolves a focused visual system, the prey evolve to disperse; whereas when the predator evolves a broad visual system, the prey evolve to swarm. Thus, we suggest that there is a selective advantage for predators that evolve a complex visual system capable of both narrow, high-acuity vision for tracking prey as well as broad, coarse vision for prey discovery.

\section*{Methods}

To study the coevolutionary cycle between predator and prey, we create an agent-based model in which predator and prey agents interact in a continuous two-dimensional virtual environment. Each agent is controlled by a {\em Markov Network} (MN), which is a stochastic state machine that makes control decisions based on a combination of sensory inputs (i.e., vision) and internal states (i.e., memory)~\citep{Edlund2011}. We coevolve the MNs of predators and prey with a genetic algorithm, selecting for MNs that exhibit behaviors that are more effective at consuming prey and surviving, respectively. Certain properties of the sensory and motor behavior of predators and prey are implemented as constraints that model some of the differences between predators and prey observed in nature (e.g., relative movement speed, turning agility, and, for predators, maximum consumption rate). Predator confusion, described in more detail below, is implemented as a constraint on predator perception that can be varied experimentally. The source code\footnote{Model code: https://github.com/adamilab/eos} for these experiments is available online. In the remainder of this section, we summarize the evolutionary process that enables the coevolution of predator and prey behavior and visual systems, describe the sensory-motor architecture of individual agents, then present the characteristics of the environment in which predator and prey interact. A detailed description of MNs and how they are evolved can be found in~\cite{Olson2016SelfishHerd}. 

\begin{figure}[tb]
\centering
\includegraphics[width=\HalfPage]{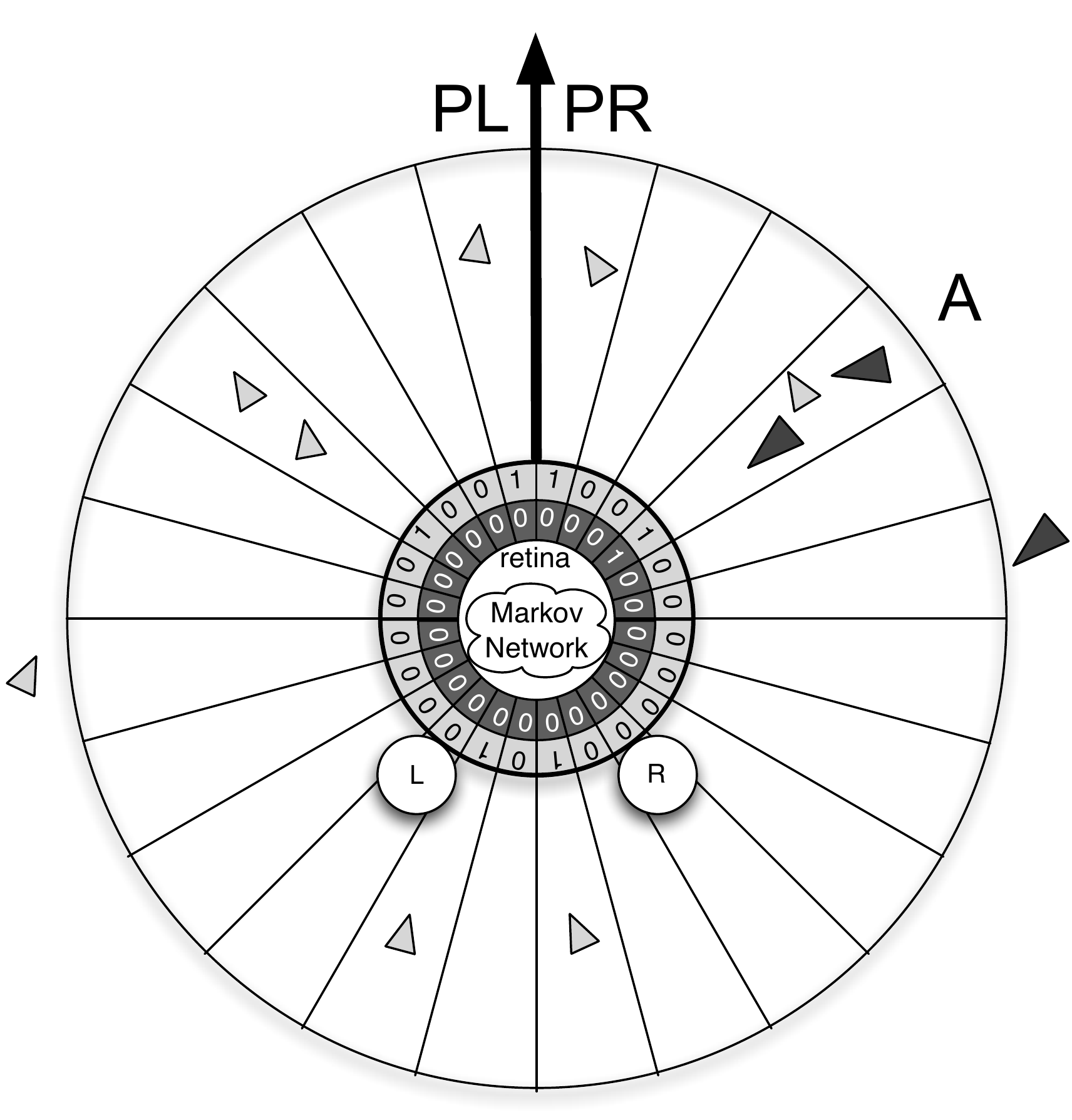}
\caption{An illustration of the predator and prey agents in the model. Light grey triangles are prey agents and the dark grey triangle is a predator agent. The prey agents have a 180$^{\circ}$ limited-distance visual system (100 virtual meters) to observe their surroundings and detect the presence of the predator and prey agents, whereas the predator agents have a variable-sized visual system that can see for 200 virtual meters. Each agent has its own Markov Network, which decides where to move next based off of a combination of sensory inputs and memory. The left and right actuators (labeled ``L" and ``R") enable the agents to move forward, left, and right in discrete steps.\label{fig:agent-illustration}}
\end{figure}

\subsection*{Coevolution of predator and prey}

We coevolve the predator and prey with a \emph{genetic algorithm} (GA), which is a computational model of evolution by natural selection~\citep{Goldberg1989}. In a GA, pools of genomes are evolved over time by evaluating the fitness of each genome at each generation and preferentially selecting those with higher fitness to populate the next generation. The genomes here are variable-length lists of integers that are translated into MNs during fitness evaluation. Furthermore, we allow the predator visual system to evolve by attaching a single integer value to each predator genome that controls the {\em predator view angle}, i.e., the size of the arc that the predator's visual system covers (see Figure~\ref{fig:agent-illustration}).


The coevolutionary process operates as follows. First, we create separate genome pools for the predator and prey genomes. Next, we evaluate the genomes' fitness by selecting pairs of predator and prey genomes at random without replacement, then place each pair into a simulation environment and evaluate them for 2,000 simulation time steps. Within this simulation environment, we generate 50 identical prey agents from the single prey genome and compete them with the single predator agent to obtain their respective fitness. This evaluation period is akin to the agents' lifespan, hence each agent has a potential lifespan of 2,000 time steps. The fitness values, calculated using the fitness function described below, are used to determine the next generation of the respective genome pools. At the end of the lifetime simulation, we assign the predator and prey genomes separate fitness values according to the fitness functions:

\begin{equation}
W_{{\rm predator}} = \sum_{t=1}^{2,000} S - A_{t}
\label{eq:pred-fitness}
\end{equation}

\begin{equation}
W_{{\rm prey}} = \sum_{t=1}^{2,000} A_{t}
\label{eq:prey-fitness}
\end{equation}
where $t$ is the current simulation time step, $S$ is the starting swarm size (here, $S$ = 50), and $A_{t}$ is the number of prey agents alive at simulation time step $t$. It can be shown that the predator fitness (Eq.~\ref{eq:pred-fitness}) is proportional to the mean kill rate $k$ (mean number of prey consumed per time step), while the prey fitness (Eq.~\ref{eq:prey-fitness}) is proportional to $(1 - k)$. Thus, predators are awarded higher fitness for capturing more prey faster, and prey in turn are rewarded for surviving longer. We only simulate a portion of the prey's lifespan where they are under predation because we are investigating swarming as a response to predation, rather than a feeding or mating behavior.

In this case, we use a GA with a population size of 100 (100 prey, 100 predators), per-gene mutation rate of 1\%, gene duplication rate of 5\%, gene deletion rate of 2\%, and mutation rate of 5\% for the predator visual system that adds a number in the range [-50$^{\circ}$, 50$^{\circ}$] to the arc size while keeping it constrained between [1$^{\circ}$, 360$^{\circ}$].

Once we evaluate all of the predator-prey genome pairs in a generation, we perform fitness-proportionate selection on the populations via a Moran process~\citep{Moran1962}, allow the selected genomes to asexually reproduce into the next generation's populations, apply random mutations to the newborn offspring, increment the generation counter, and repeat the evaluation process on the new populations until the final generation (25,000) is reached.


We perform 30 replicates of each experiment, where for each replicate we seed the prey population with a set of randomly-generated MNs and the predator population with a pre-evolved predator MN that exhibits rudimentary prey-tracking behavior with a 180$^{\circ}$ visual system. Seeding the predator population in this manner only serves to speed up the coevolutionary process, and has negligible effects on the outcome of the experiment (Figure S1 from~\citealt{Olson2013PredatorConfusion}).

\subsection*{Predator and prey agents}

Figure~\ref{fig:agent-illustration} depicts the sensory-motor architecture of predator and prey agents in this system. The retina sensors of both predator and prey agents are logically organized into ``layers", where a layer includes 12 sensors, with each sensor having a field of view of 15$^{\circ}$ and a range of 100 virtual meters.  Moreover, each layer is attuned to sensing a specific type of agent.  Specifically, the predator agents have a single-layer retina that is only capable of sensing prey.  In contrast, the prey agents have a dual-layer retina, where one layer is able to sense conspecifics, and the other senses the predator. (We note that there is only a single predator active during each simulation, hence the lack of a predator-sensing retinal layer for the predator agent.)

Regardless of the number of agents present in a single retina slice, the agents only know the agent type(s) that reside within that slice, but not how many, representing the wide, relatively coarse-grain visual systems typical in swarming birds such as Starlings~\citep{Martin1986}. For example in Figure~\ref{fig:agent-illustration}, the furthest-right retina slice has two prey in it (light grey triangles), so the prey sensor for that slice activates. Similarly, the sixth retina slice from the left has both a predator (dark grey triangle) and a prey (light grey triangle) agent in it, so both the predator and prey sensors activate and inform the MN that one or more predators \emph{and} one or more prey are currently in that slice. Lastly, since the prey near the 4th retina slice from the left is just outside the range of the retina slice, the prey sensor for that slice does not activate. We note that although the agent's sensors do not report the number of agents present in a single retina slice, this constraint does not preclude the agent's MN from evolving and making use of a counting mechanism which reports the number of agents present in a set of retina slices. Once provided with its sensory information, the prey agent chooses one of four discrete actions: (1) stay still; (2) move forward 1 unit; (3) turn left 8$^{\circ}$ while moving forward 1 unit; or (4) turn right 8$^{\circ}$ while moving forward 1 unit.

Likewise, the predator agent detects nearby prey agents using a limited-distance (200 virtual meters), segmented retina covering an evolvable angle in front of the predator that functions just like the prey agent's retina. Similar to the prey agents, predator agents make decisions about where to move next, but the predator agents move 3x faster than the prey agents and turn correspondingly slower (6$^{\circ}$ per simulation time step) due to their higher speed.

\subsection*{Simulation environment}

We use a simulation environment to evaluate the relative performance of the predator and prey agents. At the beginning of every simulation, we place a single predator agent and 50 prey agents at random locations inside a closed $512\times512$ unit two-dimensional simulation environment. Each of the 50 prey agents are controlled by clonal MNs of the particular prey MN being evaluated. We evaluate the swarm with clonal MNs to eliminate any possible effects of selection at the individual level, e.g., the ``selfish herd'' effect~\citep{Wood2007,Olson2013SelfishHerd}.


During each simulation time step, we provide all agents their sensory input, update their MN, then allow the MN to make a decision about where to move next. When the predator agent moves within 5 virtual meters of a prey agent it can see (i.e., the prey agent is anywhere within the predator's visual field), it automatically makes an attack attempt on that prey agent. If the attack attempt is successful, the target prey agent is removed from the simulation and marked as consumed. Predator agents are limited to one attack attempt every 10 simulation time steps, which is called the \emph{handling time}. The handling time represents the time it takes to consume and digest a prey after successful prey capture, or the time it takes to refocus on another prey in the case of an unsuccessful attack attempt. Shorter handling times have negligible effects on the outcome of the experiment, except for when there is no handling time at all (Figure S2 from~\cite{Olson2013PredatorConfusion}).

To investigate predator confusion as an indirect selection pressure driving the coevolution of swarming, we implement a perceptual constraint on the predator agent. When the predator confusion mechanism is active, the predator agent's chance of successfully capturing its target prey agent ($P_{\rm{capture}}$) is diminished when any prey agents near the target prey agent are visible anywhere in the predator's visual field. This perceptual constraint is similar to previous models of predator confusion based on observations from natural predator-prey systems~\citep{Jeschke2005,Jeschke2007,Ioannou2008}, where the predator's \emph{attack efficiency} (\# successful attacks / total \# attacks) is reduced when attacking swarms of higher density.

$P_{\rm{capture}}$ is determined by the equation $P_{\rm{capture}} = \frac{1}{A_{\rm{NV}}}$, where $A_{\rm{NV}}$ is the number of prey agents that are visible to the predator, i.e., anywhere in the predator agent's visual field, \emph{and} within 30 virtual meters of the target prey. By only counting prey near the target prey, this mechanism localizes the predator confusion effect to the predator's retina, and enables us to experimentally control the strength of the predator confusion effect.

\begin{figure}
\centerline{\includegraphics[width=\HalfPage]{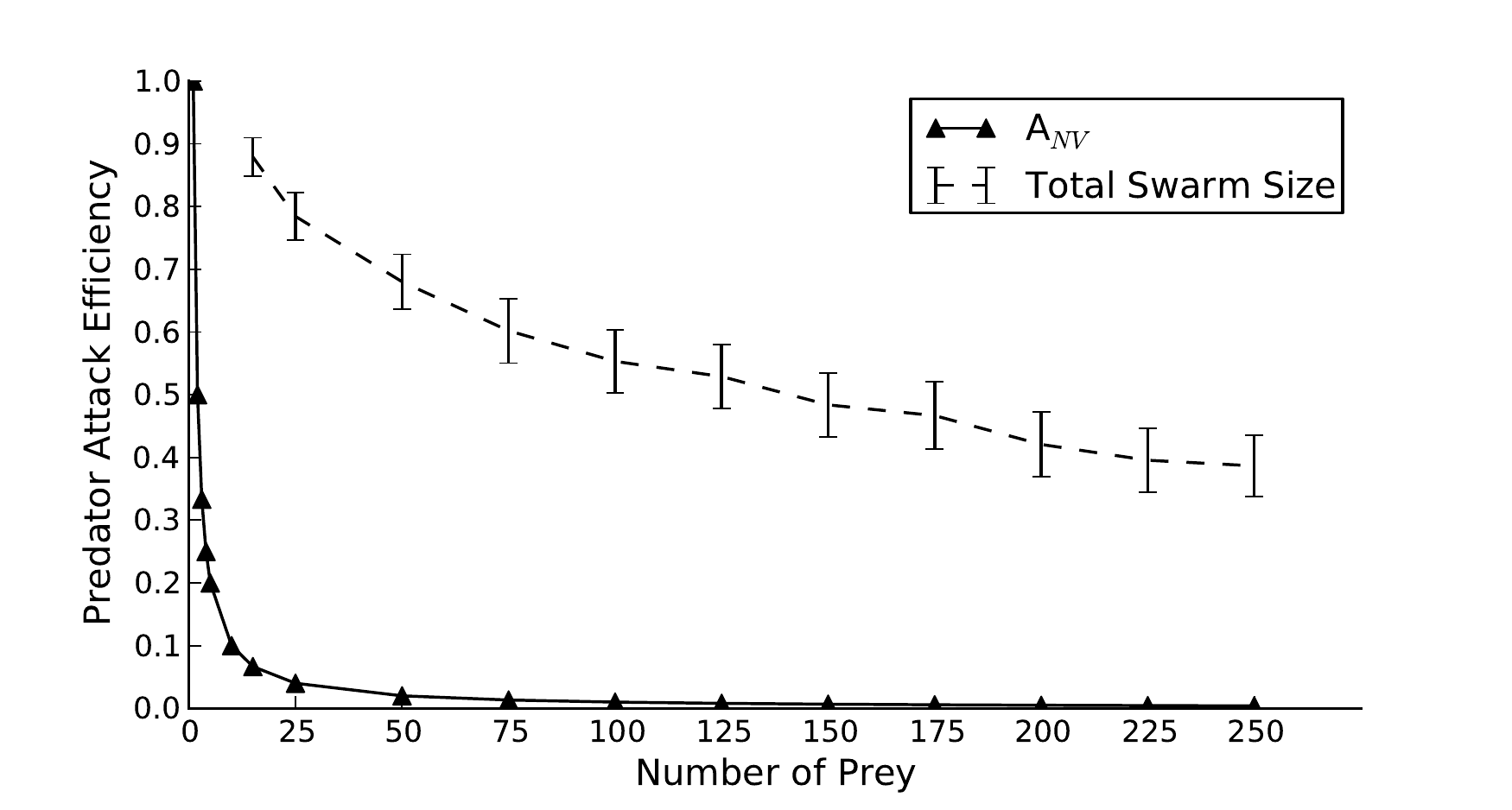}}
\caption{Relation of predator attack efficiency (\# successful attacks / total \# attacks) to number of prey. The solid line with triangles indicates predator attack efficiency as a function of the number of prey within the visual field of the predator ($A_{\rm{NV}}$). Similarly, the dashed line with error bars shows the actual predator attack efficiency given the predator attacks a group of swarming prey of a given size, using the $A_{\rm{NV}}$ curve to determine the per-attack predator attack success rate. Error bars indicate two standard errors over 100 replicate experiments.\label{fig:predator-attack-efficiency}}
\end{figure}

Although our predator confusion model is based on the predator's retina, it is functionally equivalent to previous models that are based on the total swarm size (Figure~\ref{fig:predator-attack-efficiency}, dashed line), see, e.g., ~\citep{Jeschke2005,Tosh2006,Jeschke2007,Ioannou2008}. As shown in Figure~\ref{fig:predator-attack-efficiency} (solid line with triangles), the predator has a 50\% chance of capturing a prey with one visible prey near the target prey ($A_{\rm{NV}} = 2$), a 33\% chance of capturing a prey with two visible prey near the target prey ($A_{\rm{NV}} = 3$), etc. As a consequence, prey are in principle able to exploit the combined effects of predator confusion and handling time by swarming.

\section*{Results}

To evaluate the evolved prey behavior quantitatively, we obtain the line of descent (LOD) for every replicate experiment by tracing the ancestors of the most-fit prey MNs in the final population until we reach the randomly-generated ancestral MN with which the starting population was seeded (see~\citealt{Lenski2003} for an introduction to the concept of a LOD in the context of computational evolution). For each ancestor in the LOD, we characterized the behavior with a common behavior measurement called \emph{swarm density}~\citep{Huepe2008}. We measured the swarm density as the mean number of prey within 30 virtual meters of each other over a lifespan of 2,000 simulation time steps, which provides an indication of how closely the prey are staying near each other on average. Similarly, we evaluate the predator's view angle by tracing the LOD of the most-fit predator and observing the view angle of each ancestor.

In~\citealt{Olson2013PredatorConfusion}, we showed that when the predator's visual system only covered the frontal 60$^{\circ}$ or less, swarming to confuse the predator was no longer a viable adaptation (as indicated by a mean swarm density of $0.68\pm0.02$ at generation 1,200). In this case, the predator had such a narrow view angle that few swarming prey were visible during an attack, which minimizes the confusion effect and correspondingly increases the predator's capture rate (Figure S8 from~\citealt{Olson2013PredatorConfusion}). When the predator's visual system was incrementally modified to cover the frontal 120$^{\circ}$ and beyond, swarming again became an effective adaptation against the predator due to the confusion effect (indicated by a mean swarm density of $6.13\pm0.76$ at generation 1,200). This suggests that the predator confusion mechanism may not only provide a selective pressure for the prey to swarm, but it could also provide a selective pressure for the predator to narrow its view angle to become less easily confused.

\begin{figure*}
\centerline{\includegraphics[width=\textwidth]{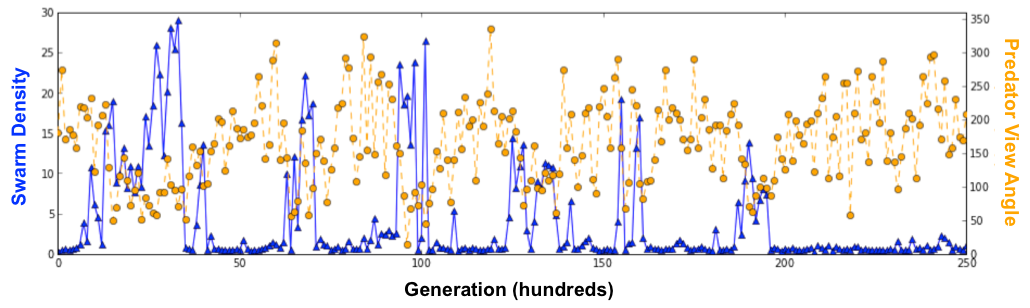}}
\caption{Swarm density and predator view angle from the LOD of a single coevolution experiment. The predator and prey populations repeatedly cycle between different states of view angles and behaviors.\label{fig:pred-prey-coev-cycle}}
\end{figure*}

\begin{figure}
\centerline{\includegraphics[width=\HalfPage]{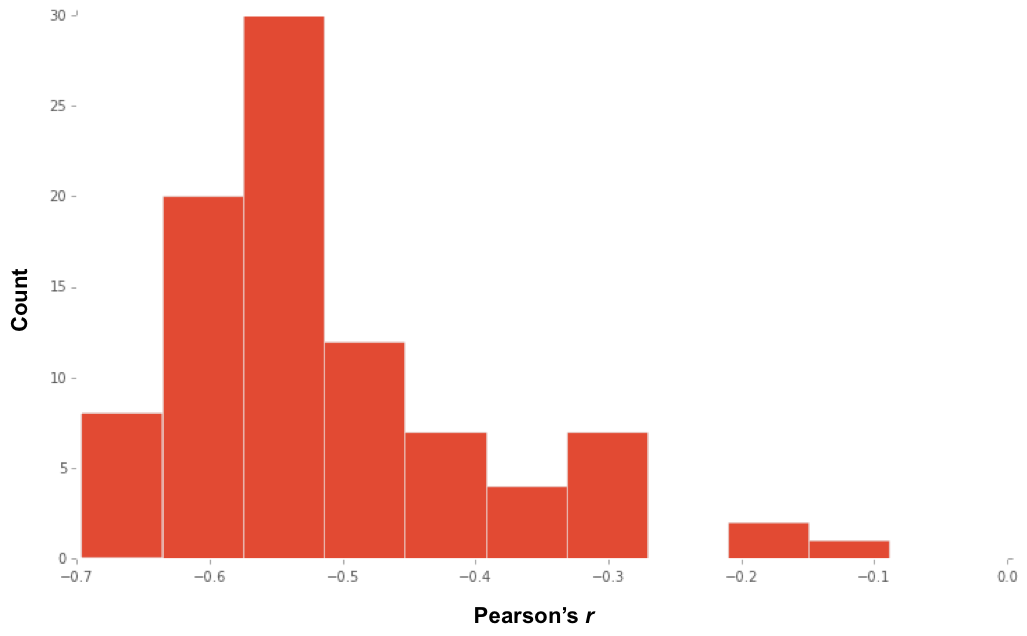}}
\caption{Pearson's \emph{r} between swarm density and predator view angle from the LODs of 30 coevolution experiments. All coevolution experiments have a negative correlation between swarm density and predator view angle, indicating that when swarm density goes up, predator view angle goes down, and vice versa. $P <= 0.001$ for all correlations.\label{fig:sdc-pva-coev-pearsons-r}}
\end{figure}

\begin{figure}
\centerline{\includegraphics[width=\HalfPage]{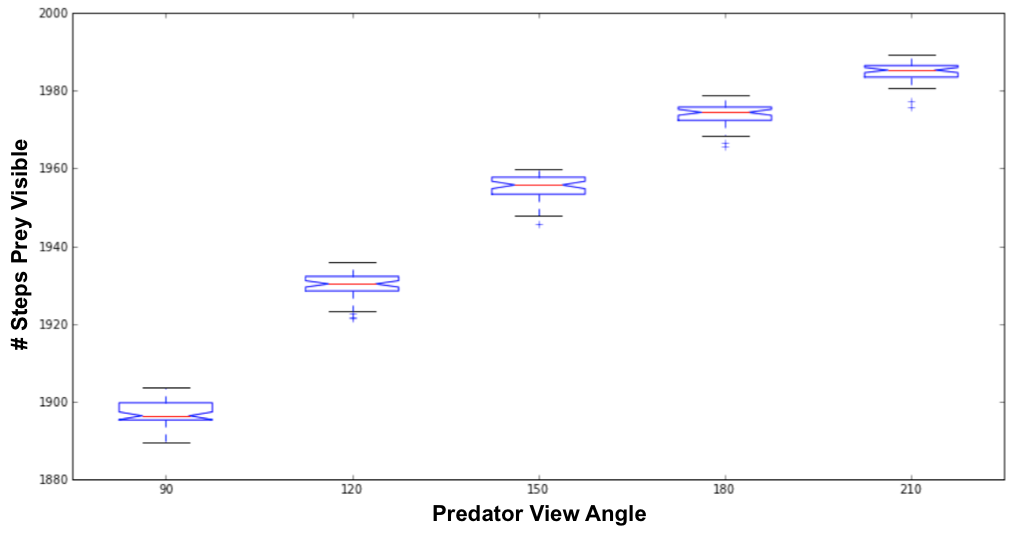}}
\caption{Number of simulation time steps that prey are present anywhere in an evolved predator's visual system depending on the predator's view angle. Each box plot represents 30 replicates, and the notches represent the 95\% confidence interval around the median. Here, the predator is competed against dispersive prey. Predators with higher view angles are more likely to have prey anywhere in their visual system at a given time. $P <= 0.001$ between all view angles with a Kruskal-Wallis multiple comparison.\label{fig:pva-efficient-hunting-visible}}
\end{figure}

\begin{figure}
\centerline{\includegraphics[width=\HalfPage]{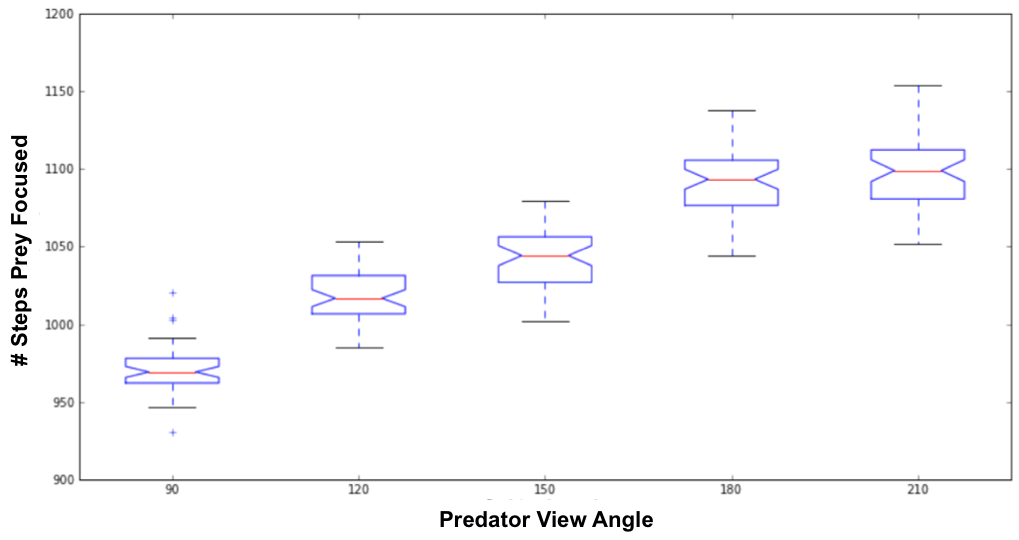}}
\caption{Number of simulation time steps that prey are visible in a portion of an evolved predator's visual system that it pays attention to, depending on the predator's view angle. Each box plot represents 30 replicates, and the notches represent the 95\% confidence interval around the median. Here, the predator is competed against dispersive prey. Predators with higher view angles are more likely to spot prey at a given time, which increases their foraging efficiency. $P <= 0.001$ between all view angles except 180 vs. 210 with a Kruskal-Wallis multiple comparison.\label{fig:pva-efficient-hunting-focused}}
\end{figure}

\begin{figure}
\centerline{\includegraphics[width=\HalfPage]{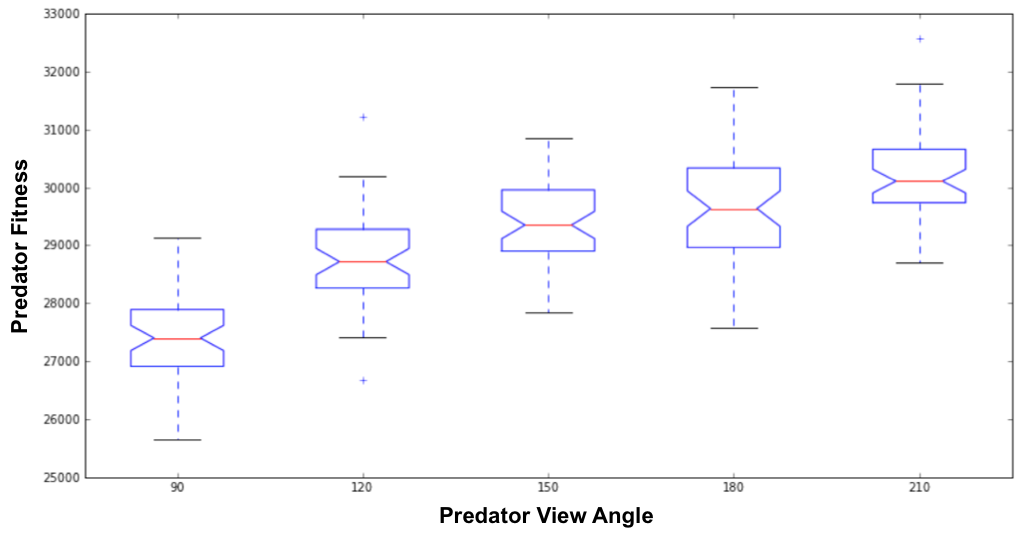}}
\caption{Fitness of an evolved predator when competed against dispersive prey, depending on the predator's view angle. Each box plot represents 30 replicates, and the notches represent the 95\% confidence interval around the median. Predators with higher view angles forage for prey more efficiently, thus capturing more prey in their lifetime and improving their fitness. $P <= 0.001$ between all view angles except 150 vs. 180 and 180 vs. 210, Kruskal-Wallis multiple comparison.\label{fig:pva-higher-fitness}}
\end{figure}

When we allow the predator view angle to coevolve along with the predator and prey behavior, we observe that the predator populations do indeed evolve focused visual systems in response to prey swarming behavior (Figure~\ref{fig:pred-prey-coev-cycle}), as indicated by the predator view angle evolving to $<100^{\circ}$ once the prey begin to swarm. Interestingly, the predator and prey populations appear to repeatedly cycle between different states of view angles and behaviors, respectively, such that there is a significant negative correlation between the predator view angle and swarm density across all 30 coevolution experiments (Figure~\ref{fig:sdc-pva-coev-pearsons-r}). This finding is surprising because the predator population should be able to effectively ``defeat'' the swarming prey population by shrinking their visual system to the point that the prey will no longer evolve to swarm. Why then would the predator population evolve to widen their visual system once the prey evolve to disperse, and allow the prey population to again evolve swarming behavior to reduce the predators' attack efficiency?

Shown in Figure~\ref{fig:pva-efficient-hunting-visible}, when predators with fixed view angles are competed against dispersive prey, the predators with broader visual systems are more likely to find a prey anywhere in their visual system at any time. Further, Figure~\ref{fig:pva-efficient-hunting-focused} demonstrates that predators with broader visual systems are also more likely to find dispersive prey in a portion of their visual system that they pay attention to, which means they spend less time searching for prey. Thus, the increased foraging efficiency that broader visual systems offer predators against dispersive prey results in higher predator fitness (Figure~\ref{fig:pva-higher-fitness}), which explains why predators evolve higher view angles in the presence of dispersive prey.

\section*{Discussion}

\begin{figure}
\centerline{\includegraphics[width=\HalfPage]{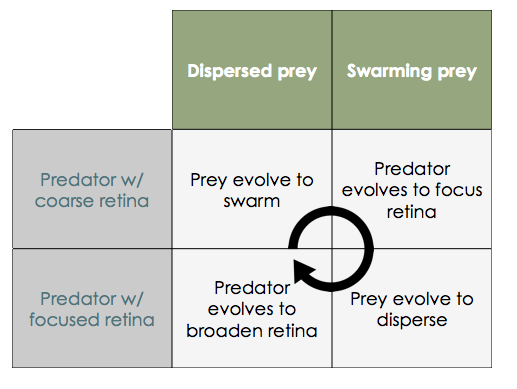}}
\caption{Diagram depicting the observed coevolutionary cycle between the predator and prey in the presence of the predator confusion effect.\label{fig:coevolutionary-cycle-diagram}}
\end{figure}

As demonstrated in Figure~\ref{fig:pred-prey-coev-cycle}, selection favors predators with a more focused visual system once swarming has evolved in prey. However, once the predators evolve a focused visual system, the prey evolve dispersive behavior in response and a coevolutionary cycle commences between the predator visual system and prey behavior. Generally, researchers assume that the evolution of social behavior is a one-way street, specifically that once social integration has arisen in evolution it may be so advantageous (compared to the cost of living in close proximity to conspecifics) that it would not be lost ~\citep{Wcislo1997}. Our findings demonstrate that at least one form of social behavior---the tendency to form cohesive swarms---can readily be gained and lost, and that the gain and loss is governed by a coevolutionary cycle that could occur between natural predators and prey due to the predator confusion effect, as depicted in Figure~\ref{fig:coevolutionary-cycle-diagram}.


Furthermore, the findings in this paper highlight a trade-off that natural predators likely experience when hunting for prey: Broader, less-focused visual systems are more useful for initially spotting prey, but focused visual systems are better adapted for tracking an individual prey down and avoiding the effects of predator confusion when hunting prey in groups. Thus, these conflicting demands imposed on the predator's visual system by the predator confusion effect could select for the evolution of complex eyes that satisfy both needs.  Indeed, many animals---both vertebrate and invertebrates---do have such complexity in the arrangement of their retinae, including the presence of a fovea in vertebrates~\citep{moore2012novel} or ``acute zones'' in invertebrates~\citep{land1997visual}. Our system could not have evolved such complexity because the retinal slices could not vary independently. 


In future work, we plan to implement a more advanced predator visual system that will allow the number of retina slices to vary, and allow each individual slice to vary in size. Through such a visual system, evolution will be capable of adjusting each retina slice as needed and allow us to explore under what conditions complex eyes will evolve. Another interesting research path would be to allow the prey visual system to coevolve as well in order to explore how evolution shapes prey visual systems in response to predation.

\section*{Conclusions}

In this paper, we implemented a computational model of evolution that allowed us to explore the coevolution of predator and prey morphology and behavior. In particular, we explored the coevolution of the predator's visual system and prey behavior and discovered that a repeated coevolutionary cycle occurs when we introduce the predator confusion effect. Furthermore, we provided evidence that the predator visual system is what drives this coevolutionary cycle, and suggested that the cycle could be closed if the predator evolves a hybrid visual system capable of narrow, high-acuity vision for tracking prey as well as broad, coarse vision for prey discovery. Thus, the conflicting demands imposed on a predator's visual system by the predator confusion effect could have led to the evolution of complex eyes in many predators.

\section*{Acknowledgements}

This research has been supported in part by the National Science Foundation (NSF) BEACON Center under Cooperative Agreement DBI-0939454. We gratefully acknowledge the support of the Michigan State University High Performance Computing Center and the Institute for Cyber-Enabled Research (iCER).

\footnotesize
\bibliography{references}
\bibliographystyle{apalike}

\end{document}